\documentclass[12pt]{article}
\usepackage{a4wide}
\usepackage{amsmath}
\usepackage{color}
\usepackage{amsopn}
\usepackage{amsfonts}
\usepackage{amssymb}
\usepackage{amsthm}
\usepackage{bbm}
\usepackage{graphicx}
\usepackage{psfrag}
\usepackage[cp1250]{inputenc}
\usepackage{pdflscape}
\usepackage[mathscr]{eucal}
\usepackage{a4}
\input xy
\xyoption{all}

\DeclareFontFamily{OT1}{rsfs}{}
\DeclareFontShape{OT1}{rsfs}{m}{n}{ <-7> rsfs5 <7-10> rsfs7 <10-> rsfs10}{}
\DeclareMathAlphabet{\mycal}{OT1}{rsfs}{m}{n}

\begin{document}

\title{\bf Arrival time in quantum mechanics \\ (demonstrated in geometrical order)}

\author{Jerzy Kijowski\\
Center for Theoretical Physics\\
Polish Academy of Sciences, Warsaw, Poland}

\date{}

\maketitle

\begin{abstract}
A geometric construction of the arrival time in conventional quantum mechanics is presented. It is based on a careful mathematical analysis of different quantization procedures for classical observables as functions of positions and momenta. A class of observables is selected which possess a unique (if any) quantized version. A simple criterion for existence of such a quantized version is formulated. These mathematical results are then applied to the classical ``arrival time'' observable.
\end{abstract}

{\em I dedicate this work to Professor Iwo Białynicki-Birula with thanks for everything I have learned from him.}

\section{Introduction}

At the turn of the 1960's and 1970's, many papers were published on the "time problem" in quantum mechanics (cf.~\cite{AB}). Numerous authors have complained that in quantum mechanics only three (among four) spacetime coordinates have quantum counterparts in the form of position operators, while the fourth, time, always remains a classical parameter of evolution. This was, according to many, a flagrant violation of the relativistic invariance that should characterize any reasonable physical theory. On the other hand, Wolfgang Pauli's comment in 1958 (\cite{Pauli}, see also paper \cite{Allcock} by G. R. Allcock) clearly indicated that treating energy (the fourth component of the "four-momentum") as momentum canonically conjugate to time and requiring these quantities to satisfy the canonical commutation rules (in order to obtain the energy-time uncertainty principle as a byproduct) leads to a contradiction with the positivity of the self-adjoint energy operator.

However, it is obvious that ``$x$'' {\em tout court} is not an observable. Measuring this quantity at different instants of times we obtain different, time-dependent results. What can be measured is ``$x(t)$'', i.e. ``the position taken by our particle at time $t$''. Similarly, the arrival time ``$t(x)$'', i.e. ``the time it takes for the particle to hit the plane $\{(x,y,z) | x = {\rm const.}\}$'', is a well-defined observable which, at least classically, can be uniquely defined and measured. For a free particle of mass $m$, whose initial position at time $t=0$ is $(x,y,z)$, this quantity is equal to:
\begin{equation}\label{arr-time}
    t(x) = - \frac {x}{v_x} = - m \cdot \frac xp \, ,
\end{equation}
where $v_x$ denotes the particle's velocity in direction of the $x$-axis, whereas $p= m v_x$ is the corresponding component of the momentum  vector. Indeed, solving equation of motion
\begin{equation}\label{e-m}
    0= x(t) = x + t v_x \, ,
\end{equation}
with respect to time we obtain \eqref{arr-time}.

According to the naive ``quantization procedure'', the quantum version ${\hat f}$ of this observable should be obtained by replacing classical position $x$ and classical momentum $p$ by the position operator $\hat x$ and the momentum operator $\hat p$:
\begin{equation}\label{q-arr-time}
    {\hat t}(x) = - m \cdot \frac {\hat x}{\hat p} \, ,
\end{equation}
or
\begin{equation}\label{q-arr-time-sym}
    {\hat t}(x) = - \frac m2 \left\{
    {\hat x}\cdot \frac 1{\hat p} +  \frac 1{\hat p} \cdot {\hat x} \right\}\, ,
\end{equation}
which looks, at the first glance, more ``hermitian''.
Unfortunately, these formulas don't make any sense: there is no self-adjoint operator that agrees with above combinations of position and momentum operators, even if restricted to a small (but dense) subspace of quantum states.

The whole "quantization procedure", i.e. representing classical observables by self-adjoint operators, goes wrong here. This is not surprising, because from a physical point of view, classical physics is the limit of quantum physics (in situations where Planck's constant is so small relative to what we can measure that it can be considered equal to zero), and not {\em vice versa}. That is, the quantum theory unambiguously implies the classical theory as an approximation. The universal validity of some ``quantization procedure'' would mean the opposite: knowing the classical theory, we would automatically know its quantum version. Such an assumption is nonsense.

However, in this article it will be proved that there is a class of observables that admit unambiguous quantization. Using these techniques, it will be shown that there is a unique way of constructing a quantum version of the arrival time, i.e. the ``arrival time operator''.

This operator was first proposed in paper \cite{JK74} (see also \cite{JK2000} and \cite{Muga}). The present author is much indebted to I. Białynicki-Birula and S. L. Woronowicz for regular discussions concerning fundamental structures of Quantum Mechanics, that we had at the beginning of 70's. These discussions were the true inspiration of the author's analysis of the problem. The construction of the operator presented in \cite{JK74} was axiomatic, based on the requirement to satisfy several, physically well-founded properties. In the present paper we present an entirely different construction, based on a mathematical analysis of the uniqueness property of the possible quantization procedures. Following the great Baruch Spinoza and his fundamental philosophical treatise {\em Ethica, ordine geometrico demonstrata}  \cite{Spinoza}, we can say that our paper describes the time of arrival ``demonstrated in geometrical order''.

The construction proposed in \cite{JK74} were later commented and criticized by many authors (see e.g. \cite{Rovelli}, \cite{Delgado}, \cite{Gianni}, \cite{Mielnik}) but none of them was able to propose another, mathematically self-consistent, construction. Nevertheless, the criticism formulated by Bogdan Mielnik and Gabino Torres-Vega in \cite{Mielnik} is well motivated from the physical point of view. It is based on the observation, that the probability that the particle hits the plane
\[
P_x :=\{ (t,x,y,z) | x = {\rm const.} \}
\]
exactly at the spacetime point $(t,y,z)$, behaves in a ``strange way'' as a function of $x$. This strange behaviour is analogous to the phenomenon known as ``probability back-flow'': even if the wave function $\psi_t(x)$ contains only the positive momenta, there might be regions, where the probability density $| \psi_t(x) |^2$ travels in negative direction as the time increases. But that's what quantum mechanics is! The superposition phenomenon, nonexisting in classical mechanics, leads inevitably to such behaviour of the probability density.

Recently, the ``time problem'' has also been intensively discussed (see e.g.~\cite{PRL}, \cite{Pullin} or \cite{tajron}). In the author's opinion, these works do not bring anything new to this discussion, because they are either mathematically inconsistent\footnote{Treating a spectral measure over a continuous spectrum as a sum over a discrete spectrum, or using "delta-normalized" wave functions, was a nice heuristic way Dirac used in 1928 to illustrate the basic concepts of quantum physics. Applying the same "techniques" in 2020 (i.e.~disregarding the 90 years of progress made here in understanding basic structures of quantum mechanics) to explain "how do we measure time in quantum theory" is unacceptable.}, or they propose new (even interesting) physical schemes, but going beyond standard quantum mechanics.

The paper is organized as follows. In Section \ref{SvH} we show how to quantize uniquely physical observables belonging to a certain, geometrically well defined class. Finally, in Section \ref{ARR}, we apply these techniques to the analysis of the observable  \eqref{arr-time} and discuss possible ways to quantize it.

\section{Schr\"odinger {\em versus} Heisenberg} \label{SvH}

The equivalence between the Heisenberg's Quantum Mechanics and the Schr\"odinger's Wave Mechanics is obvious if we substitute the self-adjoint operators ${\hat x}$ and ${\hat p} := i\hbar \cdot \frac {\partial}{\partial x}$ acting in the Hilbert space of square-integrable Schr\"odinger wave functions, for the Heisenberg q-numbers ${\hat x}$ and ${\hat p}$. However, contrary to the Heisenberg's intuition, these objects are neither finite-dimensional matrices nor continuous (bounded) operators in the infinite-dimensional Hilbert space. Consequently, even definition of their commutator
\[
    \left[ {\hat x} , {\hat p} \right] := {\hat x} \cdot {\hat p} -
    {\hat p} \cdot {\hat x} \, ,
\]
is, {\em a priori}, meaningless. The necessary, sophisticated mathematics was then elaborated by J.~v.~Neumann and his followers. In particular, the discussion of ``weak'' {\em versus} ``strong'' commutation is necessary for the uniqueness of the above Schr\"odinger representation of the Heisenberg q-numbers (i.e. uniqueness of the canonical commutation relations). Therefore, when dealing with quantum mechanics, we must remember that the algebra of unbounded operators (e.g.~their product) is an extremely subtle topic and -- if done without a proper mathematical background -- can lead to painful paradoxes.

But physically, without going deeply into this extremely difficult mathematics, the  unique representation of position and momentum as self-adjoint operators follows immediately from the Schr\"odinger equation and from the probabilistic interpretation of the wave function. The latter obviously implies the shape of the position operator:
\[
    \left( {\hat x} \psi \right) (\xi) := \xi \psi (\xi) \, .
\]
But, what is much less known, also the momentum operator is uniquely implied in wave mechanics, without any reference to Heisenberg's ``axiomatics of q-numbers''. To prove this statement, consider first a statistical ensemble of {\em classical free particles}, whose state is described by the probability density $\varphi(t;{\vec x},{\vec p})$ in the  phase space ${\cal P} = \{ ({\vec x},{\vec p} ) \}$. The corresponding densities: $\rho$ in the configuration space and $\mu$ in the momentum space are given as corresponding ``marginals'':
\begin{equation}\label{density_f}
    \rho(t;{\vec x}) := \int \varphi(t;{\vec x},{\vec p}) {\rm d}^3 p \ \ \ ; \ \ \
    \mu(t;{\vec p}) := \int \varphi(t;{\vec x},{\vec p}) {\rm d}^3 x = \mu({\vec p}) \, ,
\end{equation}
(the momentum distribution is obviously time-independent for free particles).
Knowing the dynamics of the particles (their free motion):
\[
    {\vec x}(t) = {\vec x}(0) + \frac tm {\vec p}(0) \ \ \ ; \ \ \
    {\vec p}(t) = {\vec p}(0) = {\rm const.} \, ,
\]
we know the time dependence of the density $\varphi$:
\begin{equation}\label{f(t)}
    \varphi(t;{\vec x},{\vec p}) = \varphi(0; {\vec x} - \frac tm {\vec p}, {\vec p})
\end{equation}
It is easy to check, that the momentum probability density $\mu$ can be uniquely determined {\em via} position measurements. Indeed, we have:
\begin{eqnarray*}
  \mu ({\vec p}) &=& \mu(0,{\vec p}) = \int_{\mathbb{R}^3} \varphi(0;{\vec y}, {\vec p}) {\rm d}^3y = \lim_{t \rightarrow \infty} \int_{\mathbb{R}^3} \varphi(0;{\vec y}, {\vec p} - \frac mt ({\vec y} - {\vec x}_0) {\rm d}^3y \, ,
\end{eqnarray*}
where ${\vec x}_0$ is a fixed, arbitrary point in the configuration space. Using a new variable
\[
    {\vec q} := {\vec p} - \frac mt ({\vec y} - {\vec x}_0) \Leftrightarrow
    {\vec y} := {\vec x}_0 + \frac tm ({\vec p} - {\vec q} )\ \ \ ; \ \ \
    {\rm d}^3y = \left( \frac tm \right)^3 {\rm d}^3 q
\]
we obtain:
\begin{eqnarray}
  \mu ({\vec p}) &=&  \lim_{t \rightarrow \infty} \left( \frac tm \right)^3 \int_{\mathbb{R}^3} \varphi(0;{\vec x}_0 + \frac tm ({\vec p} - {\vec q} ), {\vec q}) {\rm d}^3q \label{rho1} \\
   &=& \lim_{t \rightarrow \infty} \left( \frac tm \right)^3 \int_{\mathbb{R}^3} \varphi(t;{\vec x}_0 + \frac tm  {\vec p} , {\vec q}) {\rm d}^3q =
  \lim_{t \rightarrow \infty} \left( \frac tm \right)^3 \rho(t,{\vec x}_0 + \frac tm  {\vec p})  \, . \label{rho2}
\end{eqnarray}
Thus, by measuring the probability density $\rho$ of a particle in configuration space during its time evolution, we also obtain its probability density $\mu$ in momentum space as a result.

According to Born probabilistic interpretation of the wave function, the quantum analog of the configuration probability density is equal to
\[
\rho(t,{\vec x})= \| \psi(t,{\vec x}) \|^2 \, .
\]
Taking \eqref{rho1} -- \eqref{rho2} as the definition of momentum probability density
and using the free Schr\"odinger equation for the evolution of the wave function over time, a simple calculation is enough to prove that the above definition implies the following textbook formula:
\begin{equation}\label{mu_p}
    \mu ({\vec p}) := \lim_{t \rightarrow \infty} \left( \frac tm \right)^3
    \left\| \psi(t,{\vec x}_0 + \frac tm  {\vec p}) \right\|^2
    = \| \widetilde{\psi}(t,{\vec p}) \|^2\, ,
\end{equation}
where $\widetilde{\psi}$ denotes the Fourier transformation of $\psi$:
\[
    \widetilde{\psi} (t, {\vec p}) := \frac 1{\left( \sqrt{2\pi \hbar} \right)^3} \ \  \int_{\mathbb{R}^3} \psi(t, {\vec x})
    \exp\left( -\ i\ \frac {{\vec p} \cdot {\vec x}} \hbar \right) {\rm d}^3 x \, .
\]
Formula \eqref{mu_p} immediately implies the form of the momentum operator in the momentum representation and, consequently, also in the position representation:
\[
    \left( {\hat p} \widetilde{\psi} \right) (p) := p \widetilde{\psi} (p)
    \ \ \ \Longleftrightarrow \ \ \
    \left( {\hat p} {\psi} \right) (x) := \frac \hbar i \frac {\partial}{\partial x}
    \psi (x)
     \, ,
\]
without resorting to extremely sophisticated version of the Heisenberg's axiomatics, where a completely non-intuitive, ``strong commutation relations'' between positions and momenta must be assumed a priori. To the author's knowledge, the only textbook on quantum mechanics that derives the momentum operator in this way - the physically most intuitive - and does not postulate it {\em a priori} is the excellent book by Białynicki-Cieplak-Kamiński \cite {BCK}.

The advantage of the geometric description of quantum physics based on Schr\"odinger's wave approach over Heisenberg's algebraic formulation is particularly evident when we try to ``quantize'' more complex observables of the form $f(x,p)$. In particular, consider observables that are linear with respect to momentum:
\begin{equation}\label{lin-obs}
    f({\vec x}, {\vec p}):= X^k ({\vec x}) p_k \, ,
\end{equation}
where $X^k({\vec x})$ is an arbitrary vector field on the configuration space. Even if the quantum operators ${\hat X}^k:= X^k({\hat x})$ and ${\hat p}_k$ have already been explicitly defined, their product depends on the order of the multiplication. Unfortunately, even the symmetric order:
\begin{equation}\label{sym_f}
    {\hat f} = \frac 12 \left\{ {\hat X}^k \cdot {\hat p}_k + {\hat p}_k \cdot {\hat X}^k
    \right\} \, ,
\end{equation}
although formally ``Hermitian'', does not guarantee the self-adjointness of the resulting operator.

We are going to propose in the sequel a simple, geometric construction of the self-adjoint operator ${\hat f}$, together with a simple criterion for its existence. For this purpose observe that the vector field $X$ generates a one-parameter group ${\cal G}_t$ of local diffeomorphisms of the configuration space. These diffeomorphisms can be used to transport (drag) locally any (square-integrable) wave function. Such a transport ${\hat{\cal G}}_t$ is a unitary transformation (i.e.~does not ``lose'' any piece ``$\| \psi \|^2 {\rm d} x $'' of the particle's probability) if and only if the transformations are {\em global}, i.e.~the field $X$ is complete. But a group of unitary transformations is always of the form:
\begin{equation}\label{G_unitary}
    {\hat{\cal G}}_t = \exp \left( \frac {it}\hbar {\hat f} \right) \, ,
\end{equation}
and, whence, its {\em self-adjoint} generator ${\hat f}$ is uniquely defined. Mathematically, generator of the classical transport group is called the Lie derivative with respect to the vector field $X$ and is denoted by $\pounds_X$. We have, therefore, the following, unique formula:
\begin{equation}\label{Lie}
    \hat{f} := \frac \hbar i \pounds_X = \frac \hbar i \left(\frac {\rm d}{{\rm d}t}  {\hat{\cal G}}_t \right)_{t=0} \, ,
\end{equation}
which is automatically self-adjoint if the diffeomorphisms ${\cal G}_t$ are global. In the very special case of a constant field $X = \tfrac {\partial}{\partial x^k}$ the Lie derivative reduces to the partial derivative and, therefore, formula \eqref{Lie} reproduces the textbook formula for the momentum operator.

In this way, we get a nice and practical quantization rule for the observable $f$ together with an easy criterion for its self-adjointness, i.e.~for the reasonableness of the whole procedure. Moreover, this criterion is of topological nature, namely it imposes the existence of global solutions of of the dynamical system
\[
 \frac  {\rm d}{{\rm d}t}\ x^k(t) = X^k({\vec x}(t)) \, ,
\]
and has nothing to do with the algebraic complexity of the function $X({\vec x})$.

We will illustrate this method of ``quantization'' taking as an example a 1-dimensional problem
\[
    f(x,p) := X(x) p \, .
\]
In the simplest case, when $X(x) = 1 = {\rm const.}$, the group ${\cal G}_t$ is simply the group of translations: ${\cal G}_t(x) = x+t$. But, locally, we can always find a coordinate $s = s(x)$ such that the field $X$ is constant when expressed in terms of this coordinate, i.e.~that the following identity holds:
\begin{equation}\label{wyprost}
    X(x) \frac{\partial}{\partial x} = \frac{\partial}{\partial s} \, .
\end{equation}
To find such a new coordinate we must, therefore, solve the following differential equation:
\begin{equation}\label{ds/dx}
    \frac {{\rm d}s}{{\rm d}x} = \frac 1{X(x)} \, .
\end{equation}
If the solution is global, then $\hat{f}$ is uniquely defined as the generator of the group of translations:
\begin{equation}\label{aaa}
    \hat{f} := \frac \hbar i \frac {\rm d}{{\rm d}s} \, ,
\end{equation}
acting on wave functions in the $s$-representation. To express this operator in the original $x$-representation we must remember that the wave function {\em is not} a scalar but a ``half density''. The correct transformation formula between the two representations is implied by the following identity:
\begin{eqnarray*}
  \int \| \psi(x) \|^2 {\rm d}x  &=& \int \| \psi(x(s)) \|^2 \frac {{\rm d}x}{{\rm d}s}
  {\rm d}s =
   \int \| \psi(x(s)) \|^2 X(x(s)) \
  {\rm d}s\, .
\end{eqnarray*}
This means that the following transformation $U: L^2(\mathbb{R}) \mapsto L^2(\mathbb{R})$ between the two representations is unitary:
\begin{equation}\label{unitaryU}
        (U\psi)(s) = \psi(x(s)) \cdot \sqrt{X(x(s))} \, .
\end{equation}
The existence of the unitary operator $U$, i.e.~the global character of the group ${\cal G}_t$, is essential here. It enables us to re-calculate the Lie derivative with respect to $X$ from $x$-representation to $s$-representation and {\em vice versa}.

Geometrically, a substantial simplification of the formulae used below is obtained if we represent the quantum state by the half-density $\Psi:=\psi \cdot \sqrt{{\rm d}x}$, instead of the scalar function $\psi$. To transport such a quantity along the vector field $X$ we must transport not only the scalar factor $\psi$, but also the half-density factor $\sqrt{{\rm d}x}$. We, physicists, we perfectly know how to transport a density, like ``${\rm d}x$'', but are not used to half-densities. For this reason, we decided to use in this paper the standard, textbook notation.

In this notation we have:
\begin{equation}\label{transition_s_to_x}
        \pounds_X \psi = U^{-1}\circ \frac {\rm d}{{\rm d}s} \circ \ U \psi \, ,
\end{equation}
and, whence:
\begin{eqnarray}
  (\pounds_X \psi )(x)  &=& U^{-1}\ \frac {\rm d}{{\rm d}s}\  \left(
  \psi(x(s)) \cdot \sqrt{X(x(s))} \right) =
  U^{-1}\ X \ \frac {\rm d}{{\rm d}x}\  \left(
  \psi(x) \cdot \sqrt{X(x)} \right) \nonumber \\
   &=&  U^{-1}\left\{   \left( X^{\frac 32} \cdot \psi^\prime
   + \frac 12 \cdot \psi \cdot\ X^\prime \cdot \sqrt{X}\right)(x(s)) \right\} =
   \left( X \frac {\rm d}{{\rm d}x} + \frac 12 X^\prime \right)\psi(x) \nonumber \\
   &=& \frac 12 \left( X \frac {\rm d}{{\rm d}x}\psi + \frac {\rm d}{{\rm d}x} (X\psi)
   \right)(x) \, . \label{naive}
\end{eqnarray}
This formula for the Lie derivative, together with formula \eqref{Lie}, reproduces {\em formally} the naive quantization formula \eqref{sym_f}. Note, however, that \eqref{sym_f} does not capture the definition of the operator ${\hat f}$: indeed, this formula makes, {\em a priori}, no sense if the transport group ${\cal G}_t$, generated by $X$, is not global, or in other words, if the global unitary transformation $U$ does not exist. We conclude that a purely algebraic approach to quantization is completely inadequate, since the problem depends entirely on the analytic and topological properties of $X$.

{\bf Example 1.} For $X({\vec x})={\vec x}$ the transport group generated by $X$ is the global homothety group:
\begin{equation}\label{trans_x}
    {\cal G}_t ({\vec x}) = \exp (t) \cdot {\vec x} \, ,
\end{equation}
in the 3D case, which reduces to
\begin{equation}\label{trans_x_1D}
    {\cal G}_t (x) = \exp (t) \cdot x \, ,
\end{equation}
in the 1D case. Hence, the operators: ${\hat f} := \tfrac 12 \left( {\hat x}^k\cdot {\hat p}_k + {\hat p}_k \cdot {\hat x}^k \right)$ and its 1D analog: ${\hat f} := \tfrac 12 \left( {\hat x}\cdot {\hat p} + {\hat p} \cdot {\hat x} \right)$, are essentially self-adjoint. They are entirely described by formula \eqref{Lie} as the generators of the quantum homothety group.

In 1D case, the same result can be obtained using the variable $s$ which trivializes the field $X$ according to formula \eqref{wyprost}. For this purpose we solve \eqref{ds/dx}:
\[
    \frac {{\rm d}s}{{\rm d}x} = \frac 1x \ \ \ \Rightarrow \ \ \ s= \ln | x | \, ,
\]
and observe that translations in the variable $s$ are homotheties \eqref{trans_x_1D} in the variable $x$. Note that the Hilbert space $L^2(\mathbb{R})$ of the $x$-dependent wave functions splits naturally into the direct sum of two subspaces: $L^2(\mathbb{R}_+)$ and $L^2(\mathbb{R}_-)$, describing particles localized entirely within the positive ($\mathbb{R}_+$) and the negative ($\mathbb{R}_-$) half-axis, respectively. Each of them is isomorphic with $L^2(\mathbb{R})$-space of $s$-dependent wave functions. A homothety \eqref{trans_x} is equivalent with a simultaneous shift in the variable $s$ (i.e.~${\cal G}_t(s) := s+t$) in both subspaces.

A 3D analog of the above construction is obtained if we use spherical coordinates $(r, \vartheta , \varphi)$ and put $s = \ln r$.

{\bf Example 2.} Consider $X(x)=x^2$. To find the transport group ${\cal G}_t $ we must solve the differential equation:
\[
    \frac {\rm d}{{\rm d}t}\ x(t) = x^2 (t)\ \ \ \Rightarrow \ \ \  \frac {{\rm d}x}{x^2} = {\rm d} t \ \ \ \Rightarrow \ \ \ t + c = - \frac 1x \ \ \ \Rightarrow \ \ \
    x(t) = - \frac 1 {t+c}
    \, ,
\]
which describes all the trajectories of the field, starting from different points.
The point ${\cal G}_t (x_0) $ is defined by the initial condition
\[
   {\cal G}_0 (x_0) = x_0 \ \ \ \Rightarrow \ \ \ c = - \frac 1 {x_0} \, .
\]
This implies:
\begin{equation}\label{G_x2}
    {\cal G}_t (x) = - \frac 1 {t- \frac 1x} = \frac x{1 - tx} \, .
\end{equation}
This is not a global diffeomorphism and, whence, does not define the unitary transformation of wave functions. Consider first the case $t>0$. We see that ${\cal G}_t (x)$ is defined for $x<\tfrac 1t$ only, because it escapes to infinity as $x$ approaches the value $x=\tfrac 1t$. At the same time a substantial part of the negative half-axis, namely the half-axis $]-\infty, - \tfrac 1t[$, is not covered at all, because of the inequality ``$1-tx > -tx$'', which implies immediately
\[
    {\cal G}_t (x) = \frac x {1-tx} > - \frac 1t \, .
\]
Consequently, formula \eqref{Lie} does not define any self-adjoint operator, even if purely local considerations lead to formula \eqref{naive}. This proves that the algebraically defined operator \eqref{sym_f} is not essentially self-adjoint and, therefore, does not represent any physical observable.

Physically, the above phenomenon means that: 1) we lose a part of probability, carried by the wave function $\psi(x)$ for $x>\tfrac 1t$ and: 2) an information gap is created concerning a part of probability described by the transported wave function $\psi(x(t))$ for $x(t)< - \tfrac 1t$.

Similarly, for $t<0$: 1) we lose a part of probability, carried by the wave function $\psi(x)$ for $x<-\tfrac 1t$ and: 2) an information gap is created concerning a part of probability described by the transported wave function $\psi(x(t))$ for $x(t) > -\tfrac 1t$.

Since both parts: 1) information loss and 2) information gap fit together perfectly, we can use the first one to plug the second one. Mathematically, this means that we can treat transformation \eqref{G_x2} as a global, measurable isomorphism of the real line $\mathbb{R}$. Its singularity at the single point $x=\tfrac 1t$ does not produce any problem (the transformation, even if non-continuous, is still measurable and invertible). When used to transport wave functions, it defines a continuous group of unitary transformations. Its generator \eqref{Lie} is, therefore, a self-adjoint extension of naively defined operator \eqref{sym_f}.

Physically, however, the original disease of \eqref{sym_f} has not been cured, because the above ``plugging procedure'' is not unique. Indeed, the information loss can be plugged into the information gap with an arbitrary, constant phase change ``$\exp(i\varphi)$'', which leads to another self-adjoint extension. This means that in this case operator \eqref{sym_f} has many inequivalent self-adjoint extensions and, physically, this formula is meaningless.

{\bf Example 3.} For $X(x) = x^3$ we obtain
\[
     {\cal G}_t(x) = \frac x{\sqrt{1 - tx^2}} \, ,
\]
which, when used to transport wave functions, would imply that there is an information loss for $|x| > \frac 1{\sqrt{t}}$ and no information gap to plug it into. There is no way to repair this disease, and we conclude that formula \eqref{sym_f} does not define in this case any physical observable.

To conclude these technical remarks about quantization, we stress that the canonical transformation
\[
    (x,p) \mapsto (p, -x) \, ,
\]
enables us to quantize in a similar way functions which are linear in the position variable: ``$f({\vec x},{\vec p}):= X_k({\vec p}) \cdot x^k$''. Indeed, the quantity $X$ defines a vector field on the space of momenta and can, therefore, be used to transport wave functions in momentum representation. Hence, the whole construction presented above applies here.

\section{Arrival time} \label{ARR}
The arrival time \eqref{arr-time} can thus be considered as a vector field on the space of momenta
\begin{equation}\label{arr-field}
    X := X(p) \frac {\partial} {\partial p} \ \ \ {\rm where} \ \ \
    X(p) = \frac mp \, ,
\end{equation}
(remember that the momentum canonically conjugate to $p$ is equal to ``$-x$'' and not ``$x$''). This vector field can be used to transport wave functions in the momentum representation. To find the result of such a transport let us ``straighten'' this fields similarly as in formula \eqref{wyprost}, i.e.~let us find a new variable $s = s(p)$ such, that the field is constant with respect to this variable:
\begin{equation}\label{wyprost_p}
    X(p) \frac {\partial} {\partial p} = \frac {\partial} {\partial s} \, .
\end{equation}
To find such a new coordinate we must, therefore, solve the following differential equation:
\begin{equation}\label{ds/dx_1}
    \frac {{\rm d}s}{{\rm d}p} = \frac 1{X(p)} = \frac pm \ \ \ \Longrightarrow \ \ \
     s(p) = \tfrac {p^2}{2m} = E_{kinetic}   \, ,
\end{equation}
(a possible additive constant is irrelevant here). Unfortunately, this is is not a global coordinate on the real axis representing all possible values of the momentum $p$. The two half-axes $\mathbb{R}_+$ and $\mathbb{R}_-$ in the momentum representation are covered by two identical copies of the positive half-axis $\mathbb{R}_+$ in the ``energy representation'' (or the $s$-representation). The field \eqref{wyprost_p} acts on both half-axes independently and is not complete on both of them. There is no way to cure this disease. The corresponding ``momentum operator on a half-line'' defined as ``$\tfrac \hbar i \tfrac {\partial} {\partial s}$'' has no self-adjoint extension. In other words: algebraic formule \eqref{q-arr-time} and \eqref{q-arr-time-sym} do not define anything, which could define a reasonable physical observable.  This statement can be treated as an independent proof of the W. Pauli's statement \cite{Pauli}, that there is no quantum observable corresponding to the classical arrival time \eqref{arr-time}.

In paper \cite{JK74} I have proposed to replace $t$ by another observable, namely:
\begin{equation}\label{Arrival}
    T =  - m \cdot \frac x{|p|} =  {\rm sgn}(p) t  \, ,
\end{equation}
where the symbol ``${\rm sgn}(p)$'' represents the ``sign function'': it takes value $+1$ for $p>0$ and $-1$ for $p<0$. This observable can be called the ``oriented arrival time'': it reproduces arrival time for ``right movers'' and ``minus arrival time'' for ``left movers''. The corresponding vector field to quantize is now
\[
   X := X(p) \frac {\partial} {\partial p} \ \ \ {\rm where} \ \ \
    X(p) = \frac m{|p|} \, .
\]
Consequently, equation \eqref{ds/dx_1} is replaced by:
\begin{equation}\label{ds/dx_oriented}
    \frac {{\rm d}s}{{\rm d}p} = \frac {|p|}m \, ,
\end{equation}
and the corresponding variable $s(p)$
\begin{equation}\label{arr-vect-4}
    s = \mbox{\rm sgn}(p) \cdot  E_{kinetic} = \left\{
    \begin{array}{r} \frac  {p^2}{2 m} \ \ \mbox{\rm for} \ \ p > 0 \, , \\
                                              \\
     -\frac  {p^2}{2 m} \ \ \mbox{\rm for} \ \ p < 0 \, ,
     \end{array}
     \right.
\end{equation}
is global. Hence, the unitary operator \eqref{unitaryU} does exist. Consequently, the field $X = \tfrac {\partial} {\partial s}$ is perfectly complete and defines uniquely the self-adjoint operator ${\hat T}$ as the generator \eqref{Lie} of the translation group in variable $s$. Formula \eqref{transition_s_to_x} enables for the transition from the ``oriented energy'' or $s$-representation to the momentum or p-representation.

According to \eqref{unitaryU}, quantum states are described in $s$-representation by the following wave functions:
\begin{equation}\label{wf_phi}
    \widetilde{\phi} (s) := (U\widetilde{\psi})(s) = \widetilde{\psi}\left({\rm sgn}(s) \sqrt{2m|s|}\right)\cdot
    \sqrt{\frac m {\sqrt{2m|s|}}} =
    \widetilde{\psi}\left({\rm sgn}(s) \sqrt{2m|s|}\right)\cdot
    \sqrt[4]{\frac m {2|s|}}
    \, ,
\end{equation}
where $\widetilde{\psi}(p)$ is the standard wave function in the momentum representation. Moreover, according to \eqref{arr-vect-4}, the variable $p$ was replaced by:
\[
    p = {\rm sgn}(s) \sqrt{2m|s|} \, .
\]
The transformation $\widetilde{\psi} \mapsto \widetilde{\phi}$ is, indeed, unitary because we have:
\begin{eqnarray*}
  \left| \widetilde{\phi}(s) \right|^2 {\rm d}s &=& \left| \widetilde{\psi}(p) \right|^2
  \sqrt{\frac m {2|s|}}{\rm d}s =
  \left| \widetilde{\psi}(p) \right|^2 \sqrt{\frac {m^2} {p^2}}{\rm d}s
  = \left| \widetilde{\psi}(p) \right|^2 {\rm d}p \, .
\end{eqnarray*}

According to formula \eqref{aaa}, operator $\hat T$ in the $s$-representation is defined as
\begin{equation}\label{aab}
    \hat{T} := \frac \hbar i \frac {\rm d}{{\rm d}s} \, .
\end{equation}
This means that the (inverse) Fourier transformation $\phi (T)$ of the function $\widetilde{\phi}$:
\[
    \phi (t, T) := \frac 1{\left( \sqrt{2\pi \hbar} \right)^3} \ \  \int_{\mathbb{R}^3} \widetilde{\phi}(t, s)
    \exp\left( \ i\ \frac {s \cdot T} \hbar \right) {\rm d}s \, .
\]
describes the spectral resolution of this operator. Physically, this means that that the probability that the measurement of the observable $\hat{T}$ gives a result $T \in [a,b] \subset \mathbb{R}$ is equal to
\begin{equation}\label{prob_[a,b]}
    P(T \in [a,b]) = \int_a^b \left| {\phi}(T) \right|^2 {\rm d}T \, .
\end{equation}
If a particle beam contains {\em a priori} only ``right-movers'', without any contribution from ``left-movers'', then both arrival times (the oriented and non-oriented ones) can be identified ($T=t$). Hence, the above probability density correctly describes the arrival time and properly implements Allcock's idea regarding time measurements in Quantum Mechanics. Moreover, the Sch\"odinger evolution of the wave function is especially simple in this representation because it is given by the time translation $T \rightarrow T+t$.

Also for a beam containing ``left movers'' exclusively, the value \eqref{prob_[a,b]} has a clear physical interpretation: it represents the probability that the measured arrival time will belong to the interval $[-b, -a]$, i.e.~the physical arrival time coincides with $-T$.

For an arbitrary wave function, we can always decompose the quantum state $\psi$ into the superposition $\psi = \psi_+ + \psi_-$, where $\psi_+$ represents the right moving component and $\psi_-$ represents the left moving component. In the momentum representation this decomposition is obvious:
\begin{equation}\label{l-r docomp}
    \widetilde{\psi}_+ (p) = \left\{
    \begin{array}{r}
    \widetilde{\psi}(p) \ \ \mbox{\rm for} \ \ p > 0 \, , \\
                                              \\
     0 \ \ \mbox{\rm for} \ \ p < 0 \, ,
     \end{array}
     \right. \ \ \ ; \ \ \
     \widetilde{\psi}_- (p) = \left\{
    \begin{array}{r}
    0 \ \ \mbox{\rm for} \ \ p > 0 \, , \\
                                              \\
     \widetilde{\psi}(p) \ \ \mbox{\rm for} \ \ p < 0 \, ,
     \end{array}
     \right.
\end{equation}
Consequently, we have the corresponding decomposition in $s$-representation:
\[
    \widetilde{\phi} (s) = \widetilde{\phi}_+ (s) + \widetilde{\phi}_- (s) \, .
\]
Now, densities $\left| {\phi}_+ (T) \right|^2 {\rm d}T$ and $\left| {\phi}_- (T) \right|^2 {\rm d}T$ represent probability density of arrival time for right-movers and left-movers, separately.

Under Sch\"odinger evolution the component $\widetilde{\phi}_+$ travels forward in time, whereas $\widetilde{\phi}_-$ travels backward in time, and so do both probability densities: $\left| {\phi}_+ \right|^2$ and $\left| {\phi}_+ \right|^2$.  Unfortunately, there is no ``superselection rule'' between both components (right movers and right movers) of the particle beam and, whence, the total density
\[
    \left| {\phi} (T) \right|^2 = \left| {\phi}_+ (T) \right|^2 + \left| {\phi}_- (T) \right|^2  + 2 \,{\rm Re} \left(  {\phi}_+ (T)  \cdot  \overline{{\phi}_- (T)} \right)
\]
contains also the last term describing the quantum interaction between the two beams. In other words, the total probability that the particle hits the surface $x={\rm const.}$ from the left, and the total probability that the particle hits the surface $x={\rm const.}$ from the right, do not sum up to one
\[
    \int_{-\infty}^{+\infty} \left| {\phi}_+(T) \right|^2 {\rm d}T +
    \int_{-\infty}^{+\infty} \left| {\phi}_-(T) \right|^2 {\rm d}T \ne
    \int_{-\infty}^{+\infty} \left| {\phi}(T) \right|^2 {\rm d}T = 1 \, .
\]
This is because, when measuring the time of arrival, there are events that do not belong to either the first category (right-movers) or the second category (left-movers).

The complete 3D description of the arrival time requires also remaining 2 coordinates, namely $(y,z)$, as independent variables of the function $\psi$. This way both $\psi_+$ and $\psi_-$ are functions of four variables: $(x;t,y,z)$, but quantum interpretation applies to the last three only, whereas $x$ remains the purely classical parameter numbering different 3D hypersurfaces $\{ (t,x,y,z) | x = {\rm const.} \}$ in the 4D spacetime (similarly, as $t$ remains the purely classical parameter of the wave function $\psi = \psi(t;x,y,z)$ in the position representation).

\section{Conclusions}

The result presented in paper \cite{JK74}, and then simplified slightly in \cite{JK2000}, was obtained in an axiomatic way. The probability density \eqref{prob_[a,b]} was derived as a unique quantity satisfying several physically motivated axioms. Such a derivation is similar to the construction of the Newton--Wigner position operator (see \cite{NW}) in relativistic quantum mechanics, where the ``up-movers'' (i.e.~particles) and the ``down-movers'' (i.e.~antiparticles) were also treated separately (see also \cite{gerd}) and every quantum state can be understood as a superposition of two components.

The author emphasizes that the techniques used here are based on the geometrical interpretation of the wave function as a half-density defined in the configuration space of the particle. Such an interpretation follows directly from Schr\"odinger's formulation of ``wave mechanics''. This formulation also contains the possibility of giving meaning to Heisenberg's purely algebraic formulation which, contrary to popular creeds, is not equivalent to the former. Indeed, in order to make sense of Heisenberg's formulation, one must first answer two questions: 1) ``What are those ``q-numbers'' (correct answer: ``non-bounded operators in a Hilbert space''), and then: 2) ``How the commutator of non-bounded operators is defined'' (correct answer: ``in the so-called strong sense''). Without these two steps - highly non-intuitive from the point of view of physics - the entire Heisenberg axiomatics does not make sense, and its computational possibilities do not extend beyond the (linear!) harmonic oscillator.

As a mathematical curiosity, it is worthwhile to notice that the observables ``at most linear in $p$'' and ``at most linear in $x$'' span (in a certain, mathematically well-defined sense) the space of all observables $f(x,p)$. Quantization of $f$ based on its approximation by functions belonging to those two categories for which the quantization rule is unique implies the unique quantization rule for $f$. It turns out that this rule coincides with the classical Weyl rule (see e.g.~\cite{Weyl}).

\section*{Acknowledgements}

This research was partially supported by Narodowe Centrum Nauki (Poland) under Grant No. 2016/21/B/ST1/00940.


\begin{thebibliography}{666}

\bibitem{AB} Y.~Aharonov, D.~Bohm, {\it Time in the Quantum Theory
and the Uncertainty Relation for Time and Energy} Phys. Rev. {\bf 122}
(1961) p. 1649

\bibitem{Pauli} W.~Pauli {\em Die allgemeinen Prinzipien der Wellenmechanik}, in S.~Flugge (Ed.) {\em Encyclopedia of Physics}, vol. {\bf 5/1}, Springer, Berlin (1958), see footnote on p. 60.

\bibitem{Allcock} G.~R.~Allcock, {\em The Time of Arrival in Quantum Mechanics}, Ann. Phys. (N.Y.) {\bf 53} (1969), I p. 253 -- 285, II p. 286 -- 310 and III p. 311 -- 348.


\bibitem{JK74} J.~Kijowski, {\em On the time operator in quantum mechanics and the
Heisenberg uncertainty relation for energy and time}, Rep. Math.
Phys. {\bf 6} (1974) p. 361 -- 386.

\bibitem{JK2000}  J.~Kijowski, {\em Comment on the ``arrival time'' in quantum mechanics},
Phys. Rev. {\bf A 59} (1999) p. 897 -- 899. DOI:https://doi.org/10.1103/PhysRevA.59.897

\bibitem{Muga} J.~G.~Muga, C.~R.~Leavens, {\em Arrival time in quantum mechanics}, Physics Reports, vol. {\bf 338}, n. 4 (2000). DOI:https://doi.org/10.1016/S0370-1573(00)00047-8

\bibitem{Spinoza} Baruch Spinoza, {\em Ethica, ordine geometrico demonstrata}, \\ $https://en.wikipedia.org/wiki/Ethics\_(Spinoza\_book)$

\bibitem{Rovelli} N.~Grot, C.~Rovelli, R.~S.~Tate, {\em Time of arrival in quantum mechanics}, Phys. Rev. {\bf A 54} (1996) p. 4676 -- 4690. DOI:https://doi.org/10.1103/PhysRevA.54.4676

\bibitem{Delgado} V.~Delgado, J.~G.~Muga, {\it Arrival time in quantum
mechanics}, Phys. Rev. {\bf A 56} (1997) p. 3425-3435. DOI:https://doi.org/10.1103/PhysRevA.56.3425

\bibitem{Gianni} R.~Giannitrapani, {\it On the Time Observable in Quantum
Mechanics}, Int. Journ. Theor. Phys. {\bf 36} (1997) p. 1601

\bibitem{Mielnik} B.~Mielnik, G.~Torres-Vega, {\em “Time operator”:~the challenge
persists}, Concepts of Physics, Vol. II (2005) p. 81-102. DOI:10.48550/arXiv.1112.4198



\bibitem{PRL} L.~Maccone, K.~Sacha, {\em Quantum Measurements of Time}, Phys. Rev. Lett. {\bf 124}, (2020) 110402. DOI:https://doi.org/10.1103/PhysRevLett.124.110402

\bibitem{Pullin} R.~Gambini, J.~Pullin, {\em The solution to the problem of time in quantum gravity also solves the time of arrival problem in quantum mechanics} New J. Phys. {\bf 24} (2022) 053011. DOI:10.1088/1367-2630/ac6768

\bibitem{tajron} T.~Juri\'c , H.~Nikoli\'c, {\em Arrival time from the general theory of quantum time distributions} Eur.~Phys.~J.~Plus (2022) 137:631. DOI:10.1140/epjp/s13360-022-02854-w


\bibitem{BCK} I.~Białynicki-Birula, M.~Cieplak and J.~Kami\'nski, {\em Theory of Quanta},
Oxford University Press (1992) ISBN	0195071573, 9780195071573. Original version in Polish:  {\em Teoria Kwantów. Mechanika Falowa}, PWN, Warszawa (1991)

\bibitem{NW} T.~D.~Newton,  E.~P.~Wigner, {\em Localized States for Elementary Systems}, Reviews of Modern Physics, {\bf 21} (1949) p.~400–406.



\bibitem{gerd} J.~Kijowski, G.~Rudolph, {\em On the localization problem
in relativistic quantum mechanics}, Bull. Acad. Polon. Sci. (math.,
phys., astr.) {\bf 24} (1976) p.~1041-1048.

\bibitem{Weyl} H.~Weyl, {\em Quantenmechanik und Gruppentheorie}, Zeitschrift für Physik, {\bf 46} (1927) 1–46.



\end{thebibliography}
\end{document}